\newcommand{\beq}{\begin{equation}}
\newcommand{\eeq}{\end{equation}}
\newcommand{\bea}{\begin{eqnarray}}
\newcommand{\eea}{\end{eqnarray}}

\newcommand{\gsim}{\lower.7ex\hbox{$\;\stackrel{\textstyle>}{\sim}\;$}}
\newcommand{\lsim}{\lower.7ex\hbox{$\;\stackrel{\textstyle<}{\sim}\;$}}

\documentclass[aps,prev,preprintnumbers,floatfix,nofootinbib]{revtex4-1}
\usepackage{graphicx}
\usepackage{epstopdf}
\usepackage{mathrsfs}
\usepackage{amssymb}
\usepackage{verbatim}
\usepackage{color}
\usepackage{multirow}
\usepackage{amsmath}
\usepackage{subfig}
 \usepackage{slashed} 
\usepackage{float}
\usepackage[normalem]{ulem}

\def\stacksymbols #1#2#3#4{\def\theguybelow{#2}
    \def\vp{\lower#3pt}
    \def\sp{\baselineskip0pt\lineskip#4pt}
    \mathrel{\mathpalette\intermediary#1}}

\def\intermediary#1#2{\vp\vbox{\sp
     \everycr={}\tabskip0pt
     \halign{$\mathsurround0pt#1\hfil##\hfil$\crcr#2\crcr
              \theguybelow\crcr}}}

\def\be{\begin{equation}}
\def\ee{\end{equation}}
\def\bea{\begin{eqnarray}}
\def\eea{\end{eqnarray}}

\def\sp{\;\;\;,\;\;\;}

\def\lsim{\raise0.3ex\hbox{$\;<$\kern-0.75em\raise-1.1ex\hbox{$\sim\;$}}}
\def\gsim{\raise0.3ex\hbox{$\;>$\kern-0.75em\raise-1.1ex\hbox{$\sim\;$}}}

\def\inbar{\,\vrule height1.5ex width.4pt depth0pt}

\def\IC{\relax\hbox{$\inbar\kern-.3em{\rm C}$}}
\def\IQ{\relax\hbox{$\inbar\kern-.3em{\rm Q}$}}
\def\IR{\relax{\rm I\kern-.18em R}}
 \font\cmss=cmss10 \font\cmsss=cmss10 at 7pt
\def\IZ{\relax\ifmmode\mathchoice
 {\hbox{\cmss Z\kern-.4em Z}}{\hbox{\cmss Z\kern-.4em Z}}
 {\lower.9pt\hbox{\cmsss Z\kern-.4em Z}}
 {\lower1.2pt\hbox{\cmsss Z\kern-.4em Z}}\else{\cmss Z\kern-.4em Z}\fi}

\def\comment#1{}
\def\to{\rightarrow}

\def\u1x{U(1)_X}
\newcommand{\nc}{\newcommand}
\nc{\LL}{L}
\nc{\vv}{\tilde{v}}
\nc{\ccdot}{\!\cdot\!}
\nc{\gsm}{G_{SM}}
\nc{\vfive}{\mathbf{5}\oplus\mathbf{\overline{5}}}
\nc{\vten}{\mathbf{10}\oplus\mathbf{\overline{10}}}
\nc{\zhol}{Z^{\rm hol}}
\nc{\xfb}{\,{\rm fb}}

\begin{document}

%
%

\preprint{}
\preprint{}

\vspace*{1mm}

\title{Ultra-light scalar saving  the 3+1 neutrino scheme from the cosmological bounds}

\author{Yasaman Farzan}
\email{yasaman@theory.ipm.ac.ir}

\vspace{0.1cm}

\affiliation{
 Institute for Research in Fundamental Sciences (IPM),\\
P.O. Box 19395-5531, Tehran, Iran  \\
The Abdus Salam ICTP, Strada Costiera 11, 34151, Trieste, Italy}

\begin{abstract} 
The LSND and MiniBooNE results as well as the reactor and Gallium anomalies seem to indicate the presence 
of a sterile neutrino with a mass of $\sim 1$ eV mixed with active neutrinos. Such sterile neutrino can be produced in the early universe before the neutrino decoupling, leading to a contribution to the effective number of neutrinos ($N_{eff}$) as well as  to a contribution to the sum of neutrino masses which are
 in tension with cosmological observations. We propose a scenario to relax this tension by a Yukawa coupling of the sterile neutrinos to ultra-light scalar particles which contribute to the dark matter in the background. The coupling induces an effective mass for $\nu_s$ which prevents its production in the early universe. We discuss the implications for the upcoming KATRIN experiment and future relic neutrino search experiments such as PTOLEMY.
We also briefly comment on certain non-renormalizable forms of interaction between $\nu_s$  and the scalar and their consequences for the $\nu_s$ production in the early universe.
 
\end{abstract}

\maketitle




\setcounter{equation}{0}



\section{Introduction}
 The 3 neutrino mass and mixing scheme has been established as the standard paradigm to 
 explain the results from various solar, atmospheric, long baseline  and reactor neutrino experiments.
 However, there are a few hints that may point out the existence of a fourth sterile neutrino ($\nu_s$) with a mass of $\sim 1$ eV mixed with active neutrinos.  This is the essence of the so-called 3+1 neutrino scheme which has been  invoked to explain the LSND \cite{Aguilar:2001ty} and MiniBooNE  \cite{Aguilar-Arevalo:2018gpe} as well as the Gallium and reactor neutrino anomalies  \cite{Giunti:2012tn,Mueller:2011nm,Mention:2011rk,Huber:2011wv}.
 To explain the LSND and MiniBooNE anomalies, $\nu_\mu$ should partially convert en-route into $\nu_e$ which implies that the sterile neutrino has to be mixed with
 $\nu_e$ and $\nu_\mu$, simultaneously. From ICECUBE and  MINOS+, strong bounds are derived on the $\nu_s$ mixing with $\nu_\mu$
 shedding doubt on the 3+1 oscillation solution to the LSND and MiniBooNE anomalies \cite{Adamson:2017uda,Jones:2019nix,Gariazzo:2015rra}. However, the reactor \cite{Mueller:2011nm,Mention:2011rk,Huber:2011wv} and Gallium \cite{Giunti:2012tn} anomalies (the observation that at short baselines $P({\nu_e} \to {\nu_e}),P({\bar{\nu}_e} \to {\bar{\nu}_e})<1$) can be explained even if $\nu_s$ mixes only with $\nu_e$  so this solution is not ruled out  by the ICECUBE or MINOS+ results which are based on the $\nu_\mu \to \nu_\mu$ observation.
 A recent analysis shows that the solutions to the Gallium and reactor anomalies are compatible with each other within the 3+1 neutrino mixing scheme with  $|U_{e4}|^2\sim 
 0.01- 0.02$ \cite{Gal}. 
What makes  this possibility even more exciting is that a sterile neutrino mixed with $\nu_e$ will lead to observable kinks in the spectrum of beta decay \cite{us}. The upcoming KATRIN
 results can test the 3+1 solution to reactor and Gallium anomalies \cite{Arman}.

On the other hand, the mixing of $\nu_s$ with $\nu_a$ implies that in the early universe before neutrinos decouple from the plasma, the neutrino oscillation 
brings the sterile neutrinos to thermal equilibrium with the active ones \cite{Gariazzo:2019gyi}. This means the effective relativistic degrees of freedom
will increase by 1 unit ($N_{eff}=4$) which is disfavored by the CMB data \cite{Planck} as well as by the Big Bang Nucleosynthesis (BBN).
 Moreover, the production of $\nu_s$ with a mass of 1 eV in the early universe will violate the upper bound on the sum of neutrino masses which are derived by combining the CMB and BAO results \cite{Planck}. 
 
 To avoid the bounds from cosmology, various models for self-interaction of $\nu_s$ has been proposed \cite{Hannestad:2013ana,Chu:2018gxk,Dasgupta:2013zpn,Chu:2015ipa,Paul:2018njm}. (See, however, \cite{Mirizzi:2014ama,Cherry:2016jol,Saviano:2014esa}.)
 The essence of all these scenarios is that the self-interaction of $\nu_s$ will induce an effective mass for $\nu_s$ at $T>$MeV which will
 suppress the effective mixing and will therefore decrease  the $\nu_a$ to $\nu_s$ oscillation probability, $P(\nu_a \to \nu_s)$.
Notice that within these scenarios, the generated effective mass itself is given by the $\nu_s$ density. That is in order for the mechanism to be efficient, a nonzero $\nu_s$ density is required in the first place. This way the bound on $N_{eff}$ can be satisfied but the bound on the sum of masses cannot be avoided. A recent thorough study shows that by adding the BAO data, the self-interaction scenario of $\nu_s$ characterized by an effective four-Fermion interaction will be still ruled out by the BAO data \cite{Song:2018zyl}. However, if the scenario involves light states coupled to $\nu_4$ which open
 up the possibility of the removal of $\nu_4$ by annihilation \cite{Archidiacono:2014nda} or decay before the onset of structure formation (before the matter radiation equality) the BAO+CMB bound can be avoided, too.
An alternative remedy is the  late phase transition scenario proposed in \cite{Vecchi:2016lty}.
A non-renormalizable coupling of $\nu_s$ to background scalar is suggested in \cite{Zhao:2017wmo} and its consequences for the $\nu_s$
abundance is discussed.

Ref. \cite{Denton:2018dqq} proposes a $U(1)$ gauge model with a gauge boson of mass 10 eV coupled to $\nu_s$ as well as to
asymmetric dark matter. The coupling creates an effective mass for $\nu_s$ proportional to dark matter density which is sizable even for
vanishing $\nu_s$ density. The effective mixing at $T>$MeV will be then suppressed,  preventing the $\nu_s$ production.  Moreover, the new gauge interaction opens up the possibility of relatively fast decay of the $\nu_4$ components of the active neutrinos before the onset of structure formation. As a result, both the bound on $N_{eff}$ from CMB and BBN and the bound on the sum of neutrino masses from BAO and CMB can be satisfied.

In this letter, we propose a scenario for making the 3+1 solution to the short baseline anomalies compatible with cosmology based on a Yukawa coupling of $\nu_s$ to ultra-light real scalar which may be considered as dark matter.
In section II, we describe the scenario and demonstrate how it solves the tensions with cosmology. In section III, we discuss the implications for KATRIN  and PTOLEMY and formulate strategies to combine various observations to eventually elucidate the mechanism behind the absence of $\nu_s$
 in the early universe. Section IV summarizes our results.

\section{The model}
It is well-known that if dark matter (or a component of it) is of bosonic type, it can be as light as $\sim 10^{-21}$ eV. Despite its small mass, the 
ultra-light dark matter is considered to be cold because its production is non-thermal. These particles can be non-relativistic even at high
temperatures. Recently such dark matter has gained popularity in the literature as it has been advocated as a solution to the 
small scale structure tensions
within the WIMP scenario \cite{Hui:2016ltb}. As long as their de Broglie wavelength is larger than their average distance with each other, they can be described by a classical field. In particular, a real ultra-light scalar dark matter can be described as 
\be \phi=\frac{\sqrt{2 \rho_\phi}}{m_\phi}\cos (m_\phi t -\vec{p}_\phi\cdot \vec{x}) \ee where $|\vec{p}_\phi|\ll m_\phi$. 
For $t \gg 1/m_\phi$, $\rho_\phi$ (like in the case of other non-relativistic relics) scales as $T^3$. For $t \ll 1/m_\phi$, it can be shown that $\rho_\phi$
(the $00$ element of the energy-momentum  tensor, $T_{\mu \nu}$) is equal to minus the pressure, $-p_\phi$ ($T_{ii}$). Thus, the relation
$T^{\mu \nu}_{; \nu}=0$ or equivalently $\dot{\rho}_\phi+3 H(\rho_\phi+p_\phi)=0$ implies that for $t\ll 1/m_\phi$, $\rho_\phi$ and therefore the amplitude of $\phi$ remains constant.

There is a vast literature discussing the production of such light particles in the early universe with non-relativistic velocities; for a review see \cite{Marsh:2015xka}. In these mechanisms, $\phi$ is taken to be the phase of a complex salar, $\Phi=|\Phi| e^{i \phi /f_0}$ with a $U(1)$ symmetry similar to the axion of the Peccei-Quinn symmetry \cite{u1}. Once the symmetry becomes spontaneously broken, $\phi$ obtains a random value in the range $(-f_0 \pi , f_0 \pi)$. If the symmetry breaking takes place during the inflation, the whole patch within our horizon will have the same value of initial $\phi$ (plus small fluctuations). The topological defects between the patches of different constant $\phi$ will safely be diluted away during  inflation. It has been demonstrated in the literature that with this mechanism $\phi$  (consequently, $\rho_\phi$) can be large enough to account for all DM. As we shall see, our scenario works  even for smaller values of $\phi$. The quantum fluctations during inflation can provide the small variation in $\rho_\phi$ which provides the seeds for structure formation during the course of the history of the universe.
Such small variation of $\rho_\phi$  is irrelevant to our discussion.

Refs. \cite{Berlin:2016woy,Krnjaic:2017zlz,Brdar:2017kbt} assume a Yukawa coupling between $\phi$ and active neutrinos and discuss the implication of the oscillatory behavior of $\phi$ with time on the temporal modulation of various neutrino beams.
Ref. \cite{Farzan:2018pnk} assumes a gauge interaction between complex ultra-light scalar DM and leptons and studies its 
impact on the flavor ratios of cosmic neutrinos detected by ICECUBE.  Here, we assume a Yukawa coupling of the following form between real $\phi$
and $\nu_s$
\be \lambda \phi \nu_s^Tc \nu_s +{\rm H}.c.\label{coupling} \ee where $c$ is an asymmetric $2\times 2$ matrix with components equal to $\pm 1$.
Notice that this coupling is renormalizable and invariant under
 the SM gauge group. As long as $\phi$ is lighter than the lightest neutrino mass eigenstate,
$\phi$ remains stable and therefore a suitable dark matter candidate. Notice that we could write the interaction of type $\phi \bar{\nu}_s \nu_s$ with  similar results but to avoid adding new degrees of freedom, we stick to this Majorana form which does not require right-handed component for
$\nu_s$. 

The coupling in Eq. (\ref{coupling}) induces an effective mass for $\nu_s$ given by
\be \label{mEff} m_{eff}=\lambda \frac{\sqrt{2\rho_\phi}}{m_\phi} \cos (m_\phi t).\ee
Taking $m_\phi <5\times 10^{-17} e{\rm V}=\frac{1}{13~{\rm sec}}$, for up to after neutrino decoupling (to be precise until $T\sim 0.22~{\rm MeV} (m_\phi/(5\times 10^{-17}~{\rm eV}))^{1/2}$), $m_{eff}$ remains almost constant
and equal to 
$ m_{eff}= \lambda \sqrt{2 \rho_\phi^{int}}/m_\phi$ where $\rho_\phi^{int}$ is the value of $\rho_\phi$ at $t \ll 1/m_\phi$. Taking for example
	$\rho_\phi^{int}=\rho_{DM}^0 (0.22 ~{\rm MeV} \sqrt{m_\phi/(5\times 10^{-17}~{\rm eV})}/T^0)^3$
 (where the  $0$ superscript  denotes the values today), we find $m_{eff}=2.3\times 10^{24} ~{\rm eV}\lambda (5\times 10^{-17} ~{\rm eV}/m_\phi)^{1/4}$.
Notice that the format of the effective mass that $\nu_s$ receives is of the Lorentz invariant  Majorana type which should be summed with the
$\nu_s$ mass in vacuum ($m_{\nu_s}$) to obtain dispersion relation {\it i.e.,} $E_{\nu_s}^2-|\vec{p}_{\nu_s}|^2=(m_{eff}+m_{\nu_s})^2$. 
Using the superradiance argument, a vector dark matter with a mass of $6\times 10^{-20}-2\times 10^{-17}$ eV
is constrained \cite{Baryakhtar:2017ngi} but these bounds do not apply for the scalar dark matter. The superradiance bound from M87$^*$ rules out only the scalar dark matter of mass of $10^{-21}$ eV and lower \cite{Davoudiasl:2019nlo}.

Remember that in the case of the propagation of the active neutrinos in matter, 
the Lorentz violating effective mass of active neutrinos in medium ({\it e.g.,} $2\sqrt{2} G_F n_e \nu_e^\dagger \nu_e$) 
is added to $m_\nu^2/(2E_\nu)$ to obtain the Hamiltonian  governing the neutrino flavor evolution. 
 Here,  the Lorentz conserving $m_{eff}$ should be added to $m_{\nu_s}$ rather than to $ m_\nu^2/E_\nu$. In the presence of $m_{eff}\gg m_{\nu_s
	}$, we can write the effective active sterile mixing angle as
\begin{eqnarray} \label{effMIX} \sin 2 \theta_m|_T=\sin2\theta \frac{m_{\nu
		_s}}{m_{eff}}=\left\{  \begin{matrix} \sin 2\theta_m^{int} & {\rm at} & t\ll m_\phi^{-1} \cr \sin 2\theta_m^{int} \left(\frac{0.22~{\rm MeV}\sqrt{m_\phi/5\times 10^{-17}~{\rm eV}}}{T}\right)^{3/2}  & {\rm at} & t\gg m_\phi^{-1} 
\end{matrix} \right.,\end{eqnarray}
 where $\theta$ is the mixing angle in vacuum.
At early universe when $T>$ 1 MeV, the active neutrinos undergo  scattering off the neighboring neutrinos and electrons. Each electroweak scattering will convert them to  coherent active states without any $\nu_s$ component. To compute the oscillation probabilities, the evolution of full density matrix has to be  computed \cite{Gariazzo:2019gyi} which is beyond the scope of the present paper. However, for $\sin^2 2\theta_m^{int} \ll 1$,  a simplified estimate  can be made as  follows \cite{book}: The rate of 
$\nu_a$ to $\nu_s$ conversion, $\Gamma_{\nu_a \to \nu_s}$, can be estimated as $$\Gamma_{\nu_a \to \nu_s}=\frac{\sin^2 2 \theta_m^{int}}{4 \tau_\nu}$$ 
where $\tau_\nu^{-1}$ is the interaction rate of neutrinos $\tau_\nu^{-1}\sim  G_F^2 T^5$.  Thus, the contribution to $N_{eff}$ can be evaluated as 
$$\delta N_{eff}= \int_{T_{min}}^{T_{max}} \Gamma_{\nu_a \to \nu_s} dt= \frac{\sin^2 2\theta_m^{int}}{4} \int_{T_{min}}^{T_{max}} \frac{1}{\tau_\mu}  dt  \ ,$$
where $T_{min}$ is the neutrino decoupling temperature and $T_{max}$ is the maximum temperature for which $(\Delta m^2/T) t \stackrel{>}{\sim} 1$.  Notice that we use the fact that for $m_\phi \stackrel{<}{\sim} 5 \times 10^{-17}$ eV up until $T_{min}$, $m_{eff}$ and therefore $\sin^2 2\theta_m^{int}$ remain constant.
Within the canonical 3+1 scheme (in the limit $\sin 2\theta_m=\sin 2 \theta$), Ref.  \cite{Gariazzo:2019gyi} shows that for $\sin^2 2\theta 
\sim 4\times 10^{-4}$ and $\Delta m^2 \sim 1 ~{\rm eV}^2$, the contribution to $N_{eff}$ is reduced to 0.1. Scaling these results, we conclude
that taking $\sin^2 2 \theta_m^{int}=4\times 10^{-5}$, the contribution will be less than $O(0.01)$  and therefore negligible.
For $\Delta m^2 \sim 3 ~{\rm eV}^2$ and $|U_{e4}|^2\sim 2\times 10^{-2}$ (a typical solution to the Gallium and reactor neutrino anomalies \cite{Gal} which is consistent with the most recent DANSS and STEREO bounds \cite{Alekseev:2018efk}),  $\sin^2 2 \theta_m^{int}=4\times 10^{-5}$ can be achieved with $m_{eff}^{int}> 40 ~e$V  which for $\rho^{int}$ corresponding to $\rho_{DM}$ implies $\lambda>2 \times 10^{-23}$. That is taking $\lambda \stackrel{>}{\sim}2 \times 10^{-23} (m_{\nu_s}/1~{\rm eV})$, the bound on $N_{eff}$ can be safely relaxed  but below $T\sim 0.01$ MeV (well above the matter radiation equality era) as well as in the Milky Way, $m_{eff}$ can be neglected because $\rho_\phi$ and therefore the amplitude of $\phi$ will be suppressed.

Let us now discuss how the bounds from BAO and CMB on the sum of neutrino masses can be avoided. As we discussed, by choosing $\lambda>10^{-23}$, the density of the $\nu_s$ particles produced at $T\stackrel{>}{\sim}$ MeV can be reduced to an arbitrarily small value. The contribution of them to the sum of the neutrino masses can be 
estimated as $\delta N_{eff} m_{\nu_s}$. Thus, as long as $\delta N_{eff}\stackrel{<}{\sim}0.01$,
the contribution is well below the bound on the sum of neutrino masses,  $\sum_{\nu} m_\nu  $ \cite{Planck}.

For $t\stackrel{>}{\sim}1/m_\phi$, the $\phi$ field will start oscillating so $m_{eff}$ can be even negative. This will have a dramatic consequence:  At certain epochs, $\nu_s$ can become lighter than even $m_\phi$, opening the possibility of  decay of $\phi$ to $\nu_s$.
Even when 
$\nu_s$ becomes lighter than $\phi$, the perturbative lifetime of $\phi$ ({\it i.e.,} $4\pi/\lambda^2 m_\phi$) will be greater than the age of universe, however; as shown in \cite{Traschen:1990sw}, the $\phi$ field can convert into $\nu_s$ and $\bar{\nu}_s$ pairs through a mechanism known as parametric resonance production. During the epoch of our interest, the radiation dominates so  $\rho_\phi \ll \rho_{\nu_a}$. Thus, even  if $\phi$ completely decays into $\nu_s$, the effects of the produced $\nu_s$ on cosmological observation will be negligible. If $\phi$ completely decays, another particle should play the role of dark matter.
If instead of the renormalizable Yukawa coupling in Eq. (\ref{coupling}), we had taken a non-renormalizable  coupling of form $\phi^2 \nu_s^T c \nu_s$ as \cite{Zhao:2017wmo} or of form $i(\phi^* \partial_\mu \phi -\phi \partial_\mu \phi^*)\bar\nu\gamma^\mu \nu$ as
\cite{Farzan:2018pnk}, $m_{eff}$ could not have become negative and 
$\phi$ would remain stable.

 The non-adiabatic conversion of $\nu_1$, $\nu_2$ and $\nu_3$ to $\nu_4$ can lead to a tension with the bounds from BAO and CMB on the sum of neutrino masses. 
Fortunately, the adiabaticity  condition is preserved because at the resonance when
$m_{eff}=m_\nu \cos 2 \theta$, we can write
$$\frac{\dot{\theta}_m}{{\rm Effective ~ mass ~splitting}}|_{resonance}=\frac{\lambda \sqrt{2 \rho_\phi} \sin (m_\phi t)}{8 m_{\nu_s}^2 \sin^2 \theta}|_{resonance}
\simeq \frac{m_\phi (T/T^{int})^{3/2}}{8 m_{\nu_s}\sin \theta \sin \theta_m^{int} }\ll 1$$ where $T^{int}$ is the temperature at the time $t \sim m_\phi^{-1}$.
  The active background neutrinos at the start of the $\phi$ oscillation are mainly composed of $\tilde{\nu}_1$, $\tilde{\nu}_2$ and
$\tilde{\nu}_3$, 
 with a small contribution given by $\sin \theta_m$ of $\tilde{\nu}_4$ where ``$\sim$" emphasizes that these are the energy eigenstates inside the 
 (dark) matter medium. Since the neutrino propagation  remains adiabatic, after the amplitude of $\phi$ diminishes due to the expansion, the background will mainly consist of the vacuum mass eigenstates $\nu_1$, $\nu_2$ and $\nu_3$ with a small contribution from $\nu_4$ given by $\sin^2 \theta_m^{int}  \sim 10^{-5}$. As a result, the contribution from $\nu_4$ to $\sum m_\nu$ will be negligible so  the bounds from CMB and BAO on $\sum m_\nu$ will be satisfied. 

Let us now discuss the stability of the $\phi$ mass in the presence of the $\lambda$ coupling. This coupling is similar to the top Yukawa coupling in the SM and will similarly induce a quadratically divergent mass for $\phi$. Like the standard model, we assume that there is a yet unknown mechanism ({\it e.g.,} SUSY) which cancels this divergent contribution. Still to have a ``natural model", we should check whether the finite part of the contribution, $\lambda m_{\nu_s}/(4\pi)$ is smaller than $m_\phi$. Taking $\lambda \sim 10^{-23}$, we see that this condition is readily satisfied. At $T\sim 10-20$ MeV when $\nu_a \to \nu_s$ may start, a ``thermal" mass of $\sim  \lambda n_{\nu_s}^{1/3}/\sqrt{12}$ is induced in which 
$n_{\nu_s}$ is the number density of the produced $\nu_s$. Remembering that  $\sqrt[3]{n_{\nu_s}}\ll T\sim 10-20$ MeV, we find the contribution is much smaller than $(\lambda/10^{-23})10^{-17}$ eV which is smaller than our benchmark value for $m_\phi$. Similar consideration holds valid for $T<0.1$ MeV where $\nu_s$ can be resonantly populated. As a result the thermal stability is guaranteed. For the  larger values of $\lambda$, the stability can be jeopardized and a more careful study is required.

\section{Prospects for KATRIN and PTOLEMY}
The KArlsruhe TRItium Neutrino (KATRIN) experiment is designed to measure the neutrino mass by studying the endpoint of the spectrum of the 
emitted electron in the beta decay of Tritium. The experiment, which will soon release its first data, can be sensitive to the neutrino mass 
(or to be more precise to $m_{\nu_e}\equiv m_1|U_{e1}|^2 +m_2|U_{e2}|^2+m_3|U_{e3}|^2$ \cite{Farzan:2002zq})
down to 0.2 eV \cite{katrin}. On the other hand, in the framework of the $\Lambda$CDM and the standard model of particles (including neutrino mass) combinig the CMB and BAO \cite{Planck} implies that neutrino mass should be smaller than this threshold and KATRIN cannot therefore discern the shift of the endpoint of the spectrum.  
However as shown in 
\cite{KATRINCOSMOS,Oldengott:2019lke}, there are ways to relax the bounds from cosmology on the sum of the neutrino masses opening up the hope
for KATRIN to resolve a sizable shift of the endpoint and to measure $m_{\nu_e}$.

If $\nu_e$ has a $\nu_4$ component with a mass of $\sim 1$ eV, it will show up as a kink \cite{us,Arman,Riis:2010zm,Abada:2018qok}
in the spectrum of the emitted electron  at $ E_e =Q-m_{\nu_4}$ where $Q$ is the mass difference between the mother and daughter nuclei.
The height of the kink will be characterized by $|U_{e4}|^2$. Within the 3+1 solution to the LSND and MiniBooNE anomalies or the 3+1 solution to the reactor and Gallium anomalies, the size of the kink can be large enough to be resolved \cite{Arman}. Let us discuss the implications of KATRIN observations combined with other observations within our scenario.

If future studies establish  a deficit in the reactor $\bar\nu_e$ flux compatible with the 3+1 scheme with $\Delta m^2 \sim 1 ~e{\rm V}^2$ and $|U_{e 4}|^2\sim 0.01$ and  on the other hand if KATRIN observes a kink with the corresponding position and amplitude, this will be a strong hint in favor of the 3+1 scheme. 
There is a similar concept to detect relic neutrinos by the $\nu_e$ capture on Tritium. The PTOLEMY experiment is proposed to search for relic neutrinos invoking this concept \cite{Betti:2019ouf}. Similarly to KATRIN, it can also study the beta decay spectrum. Within the 3$\nu$ mass scheme, we expect a peak further away from the endpoint at $E_e=Q+m_{\nu_e}$ due to the $\nu_e$ capture on Tritium. (To be more precise, we expect three peaks at $Q+m_{\nu_1}$, $Q+m_{\nu_2}$ and $Q+m_{\nu_3}$ which overlap  with each other, looking like a single peak. Since within the $3\nu$ scheme, we expect $F_{\nu_1}:F_{\nu_2}:F_{\nu_3}=1:1:1$, the heights of these three overlapping peaks are given by $|U_{e1}|^2$, $|U_{e2}|^2$ and $|U_{e3}|^2$.) Within the $3+1$ scheme in addition to this peak, there will be another peak at $E_e =Q+m_{\nu_4}$ but with a height suppressed
by $|U_{e4}|^2$. 
We should however notice that in our scenario, as discussed in the previous section,
the contribution of $\nu_4$ to the background will be  
 negligible so there will be no second peak. There will be only one peak with the same  height  as it is expected  within  the standard $3\nu$ scheme.
Observation of a kink at KATRIN spectrum but absence of a second peak at PTOLEMY is a signature of this scenario.

\section{Summary}
We have proposed a scenario to make the 3+1 solution to the short baseline neutrino anomalies compatible with the cosmological observations. 
The scenario is based on a small Yukawa coupling between the sterile neutrino and ultra-light background scalar with a mass of $m_\phi<5\times 10^{-17}$ eV. This coupling will induce an effective mass for $\nu_s$ in the early universe when active neutrinos are still in thermal equilibrium with plasma, suppressing the effective active sterile mixing and therefore the $\nu_s$ production. This way the bound on $N_{eff}$ is satisfied. Below $T\sim 0.01$ MeV (and for sure at present) the effective mass induced by the coupling to the dark matter is negligible.


After the neutrino decoupling when the $\phi$ field starts oscillating, the effective mass induced for $\nu_s$ ($m_{eff}$) can become negative,
canceling the vacuum mass, $m_{\nu_s}$. At these epoches, $\phi$ may convert into $\nu_s \bar{\nu}_s$ via parametric resonance but since $\rho_\phi \ll \rho_{radiation}$, the total amount of  $\nu_s$ produced in this way will be too small to lead to a significant contribution to $N_{eff}$.

 We have discussed 
the interpretation of possible results from future observations of KATRIN and PTOLEMY within the framework of the present scenario and  compare it with the predictions of certain alternative frameworks.
Like the canonic $3+1$ scheme, we predict a kink at the KATRIN spectrum but the spectrum at PTOLEMY will be similar to that within the  3 $\nu$  scheme with only one peak.

Throughout this letter, our main focus was on the $3+1$ solution to the short baseline anomalies but these results can be applied to even the
$3+1$ solution to the ANITA events
\cite{Cherry:2018rxj} which relies on nonzero $|U_{\tau 4}|$ instead of nonzero $|U_{e 4}|$ or $|U_{\mu 4}|$.
After this work was submitted to the archive, Ref. \cite{Cline:2019seo} appeared which confirms the suppression of $\delta N_{eff}$ by the present scenario.

\begin{acknowledgments}

 This project has received funding from the European Union\'~\!s Horizon 2020 research and innovation programme under
  the Marie Sklodowska-Curie grant agreement No 674896 and No 690575. YF has received partial financial support from Saramadan 
  under contract No.  ISEF/M/98223. YF would like also to thank the  ICTP staff and the INFN node of the INVISIBLES network in Padova.
The author is also grateful to M M Sheikh-Jabbari for useful discussions and thanks A. Smirnov, P. Denton and J. Cline for fruitful remarks.
\end{acknowledgments}

\end{document}